\documentclass[]{aa}
\usepackage[varg]{txfonts}

\usepackage[colorlinks=true,urlcolor=blue,citecolor=blue,linkcolor=blue]{hyperref}

\bibpunct{(}{)}{;}{a}{}{,} % to follow the A&A style

\def\kms{\mbox{km s$^{-1}$}}

\newcommand{\dv}{\vec{\nabla}}
\newcommand{\vu}{\vec{u}}
\newcommand{\dd}{\mathrm{d}}
\newcommand{\ddt}[1]{\frac{\dd #1}{\dd t}}

\newcommand{\be}{\begin{equation}}
\newcommand{\ee}{\end{equation}}
\newcommand{\bea}{\begin{eqnarray}}
\newcommand{\eea}{\end{eqnarray}}

\begin{document}

\title{Tracing the evolution of radiation-MHD simulations of solar and stellar atmospheres  in the Lagrangrian frame}
\titlerunning{Tracing flows in the Lagrangrian frame}

\subtitle{ }
 
\author{Jorrit Leenaarts\inst{1}
 %\and Mats Carlsson\inst{2,3} }
}

\offprints{J. Leenaarts \email{jorrit.leenaarts@astro.su.se}}

\institute{
Institute for Solar Physics, Department of Astronomy, Stockholm University, AlbaNova University Centre, SE-106 91 Stockholm, Sweden
%\and
%Institute of Theoretical Astrophysics, University of Oslo, P.O. Box 1029 Blindern, N-0315 Oslo, Norway
%\and
%Rosseland Centre for Solar Physics, University of Oslo, P.O. Box 1029 Blindern, N-0315 Oslo, Norway
}

\date{Received; Accepted }

\abstract 
{Radiation-MHD simulations have become a standard tool to investigate the physics of solar and stellar atmospheres.}
{The aim of this paper to present a method that allows efficient and accurate analysis of flows in such simulations in the Lagrangian frame.}  
{This paper presents a method that allows the construction of pathlines given a seed point that can be chosen freely at any location and at any time during the simulation where the simulation state is stored. The method is based on passive tracer particles. Through injection of particles in expanding regions the occurrence of particle-free volumes is avoided, even in the case of strongly compressive flows.}  
    {The method was implemented in the solar and stellar atmosphere simulation code Bifrost. It  is efficient and accurate. As examples I present an analysis of a gas parcel in the convection zone and a particle in the solar transition region.}
     {} 
     
    
    \maketitle

%%%%%%%%%%%
\section{Introduction} \label{sec:introduction}
%%%%%%%%%%%

Radiation-MHD simulations have become a standard tool for investigating the physics of solar and stellar atmospheres
\citep[e.g.,][]{1982A&A...107....1N,2005A&A...429..335V,2011A&A...531A.154G,2012JCoPh.231..919F,2015arXiv150707999W}.
Such simulations are often analysed in the Eulerian frame, i.e., quantities are analysed on the computational grid. This is often a convenient choice because data from the simulation can be directly written to disk for further analysis.

For certain classes of problems this is not the optimal choice. If one is interested in the evolution of a given fluid parcel instead of the evolution at a given location in the simulation, then a description in the Lagrangian frame is more natural. Typical examples where a Lagrangian description is useful in the context of solar atmospheric simulations are are mass flows during magnetic flux emergence, mass loading of chromospheric fibrils, mass and energy cycling between the chromosphere and the corona, and flow acceleration during reconnection. 

Various authors have used Lagrangian tracking in a post processing step
\citep{2007A&A...467..703C,2007A&A...461.1163C,2009A&A...507..949T,2013ApJ...776L...4S,2013ApJ...771...20M,2014PASJ...66S..10W,2015ApJ...802..136L,2016ApJ...822...18N}.
Here the position of corks is computed after the simulation run from the velocity data that is typically stored at a time interval that is much larger than the time steps taken in the simulation. Such post-facto flow computations are thus inherently less accurate than tracer particles that are advected in tandem with the simulation. 

\begin{figure*}
\centering
\includegraphics[width=17cm]{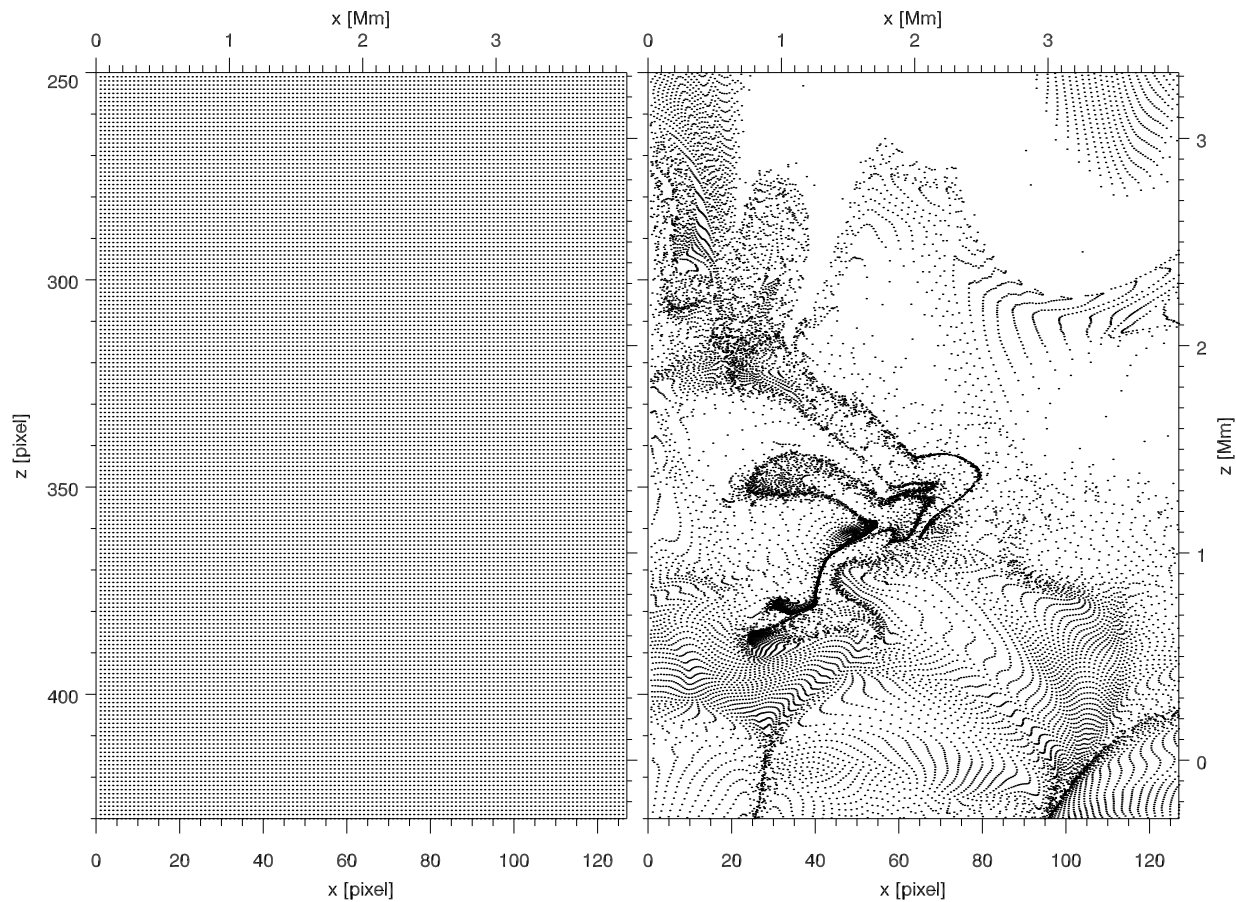}
\caption{Illustration of a highly compressive flow leading to both cork-free voids and regions with unnecessary high cork density. Left-hand panel: Initial homogeneous cork placement in a 2D simulation, with one cork per grid cell. The top of the convection zone is located roughly at $z=0$~Mm. Right-hand panel: the same simulation but now after 300~s of solar time. The bottom and left axis are in computational pixels, the upper and right axes show physical size. The panels show only a subfield of the full simulation which had $512 \times 496$ grid cells. Corks have been swept up in convective downflows and shock fronts, and large cork-free regions appear above $z=1$~Mm. }
\label{fig:void}
\end{figure*}

Flows in the convection zone and solar atmosphere are strongly compressive. If corks are injected only once and then left to evolve freely, then quickly cork-free voids will appear. Flows cannot be traced through such voids. In addition, areas with converging flows will appear where the cork density is needlessly high. I show an example in Fig.~\ref{fig:void} from a 2D radiation-MHD simulation computed with the Bifrost code 
\citep{2011A&A...531A.154G}.
It shows a small subfield from the simulation covering the very top of the convection zone and a part of the chromosphere and transition region. The initially homogenous cork distribution has become highly inhomogeneous after only 300~s of solar time. 

This is not a problem if one is interested in a specific event of short time duration in the simulation. One can then simply inject corks in a given volume and at a time just before the event of interest and rerun part of the simulation. This is however a time-consuming process, and cannot be used to study global evolution and long-term time evolution.

In this paper I present a method to analyse simulations in the Lagrangian frame based on corks that are advected during the radiation-MHD simulation. It allows tracking the history and future of the position, velocity, all forces, and all energy losses and gains of any gas parcel starting at any given time and any location in the simulation. It allows for flexible analysis of the simulations without requiring prior knowledge of the location and time of events of interest.

%%%%%%%%%%%%%%%%%%
%%%%%%%%%%%%%%%%%%

%%%%%%%%%%%%%%%%%%
\section{Methods} \label{sec:methods}
%%%%%%%%%%%%%%%%%%

%%%%%%%%%%%%%%%%%%
\subsection{The Bifrost code} 
%%%%%%%%%%%%%%%%%%

I implemented the algorithms described in Sec.~\ref{subsec:inject_remove} and~\ref{subsec:construct} in the Bifrost code
\citep{2011A&A...531A.154G}.
Bifrost solves the equations of resistive MHD on a Cartesian grid in one, two, or three dimensions. It uses a sixth-order finite difference scheme to compute spatial derivatives and a third-order predictor-corrector scheme to advance the simulation in time. Bifrost is a flexible code that can include various physical processes. In the simulations discussed here Joule dissipation, non-LTE radiative heating and cooling and thermal conduction along field lines is included. The equation of state is pre-tabulated assuming LTE for a standard solar elemental composition. For further information about Bifrost we refer to 
\citet{2011A&A...531A.154G}.
%

%%%%%%%%%%%%%%%%%%
\subsection{Pathlines and streamlines} \label{subsec:PLandSL}
%%%%%%%%%%%%%%%%%%

The corks are passively advected by the time-dependent plasma velocity field $\vu(\vec{r},t)$, i.e, the position $\vec{r}_i(t)$ of a cork labeled $i$ evolves as 
\be
\frac{\partial \vec{r}_i(t)}{\partial t} = \vu(\vec{r}_i,t), \label{eq:pathline}
\ee 
with $\vu(\vec{r}_i, t)$ the plasma flow velocity at the location of the cork. The corks thus trace out pathlines. This is in contrast to streamlines
$\vec{r}(s)$, which are defined by
\be
\frac{\partial \vec{r}}{\partial s} = \vu(\vec{r},t),
\ee 
for the path length $s$ along the streamline given the velocity field at a fixed time $t$. Pathlines and streamlines only coincide for a steady flow, a situation that in general does not hold for the solar atmosphere
\citep[e.g.,][]{2013ApJ...776L...4S}.

Numerically the velocity at the cork position is computed using fifth-order interpolation from the velocity stored on the Eulerian grid of the simulation, and the cork position is advanced in time using the same predictor-corrector scheme as used in Bifrost.

%%%%%%%%%%%%%%%%%%
\subsection{Cork injection and removal} \label{subsec:inject_remove}
%%%%%%%%%%%%%%%%%%
Most radiation-MHD  codes solve the equations on a fixed Eulerian grid. There is thus only limited flow information present on scales much smaller than a grid cell. In order to trace pathlines down to the resolution limit of the simulation one needs a cork density of order of $\sim1$ cork per grid cell. This density allows resolving pathlines down to the resolution limit, but without the overhead of multiple corks following nearly the same pathline.

In order to keep the cork density close to the desired density and avoid inhomogeneous distributions such as in Fig.~\ref{fig:void}, one needs to dynamically inject and remove corks. One sets a maximum and minimum cork density per grid cell. One also sets a control volume size. The computational domain is divided into those control volumes. After some time interval time interval $\Delta t$ the code checks the cork density in the control volumes. If the cork density is below the minimum, then new corks are injected in the control volume to raise the cork density above the minimum. If the cork density is too high, corks are removed from the control volume.

Generally, corks do not trace pathlines that span the entire simulation time, because they can be injected or removed, The core idea is to create approximate pathlines spanning the full simulation time from the cork trajectories: to do so one first selects a location and a time that provide the starting point to integrate Eq.~\ref{eq:pathline}. Then one finds the cork closest to this location, called a cork of interest (COI). The COI is followed forward in time until a) the simulation end time is reached, b) it exits the domain, or c) it is removed. If it is removed we find the next closest cork, and follow it until the simulation end time, the cork exits the domain, or is removed. We repeat this procedure until the end time of the simulation. Then we trace the cork backward in time until a) the start time of the simulation, b) the cork enters the domain, or c) it is injected. If it is injected we find the closest cork that exists in the preceding snapshot and start following that one backwards in time.

In order to be able to trace any pathline through any given point at any given time down to the resolution limit of the simulation one always needs at least one cork per grid cell.  A logical choice is therefore to set the control volume to the size of a single grid cell and the minimum cork density to one. The maximum cork density can be set to any value, but in order to keep the number of corks manageable also in large 3D simulations, a decent choice is to set the maximum cork density to two. The maximum number of corks in a simulation is then at most twice the number of grid cells.

Ideally one sets the the time interval $\Delta t$ between adjustments of the cork density to the time step of the MHD solver. This leads however to difficulties in analysing the flow paths (see Sec.~\ref{subsec:construct}) owing to the large data volumes that are generated. We therefore set  $\Delta t$ to the same time interval at which the MHD variables are written to disk, which is commonly  $\sim 10$~s. With a typical grid spacing of 20~km and flow speeds of 20~\kms\ this leads inevitably to grid cells without corks in case of divergent flows. In case flows need to be traced through such grid cells, then the time interval between injection-removal sweeps (IR-sweeps) can be decreased as much as needed, at the price of an increased amount of data to process. 

In contrast to post-processing methods, the accuracy of the individual cork trajectories themselves is not affected by the time interval at which  the IR-sweeps are performed and the data is written out. The interval only influences the time resolution with which the flows are sampled, corks are removed, and corks are injected.

Simulations of the solar atmosphere often have periodic boundary conditions in the horizontal directions, and the corks follow this behaviour. Corks that exit the computational domain on the upper and lower boundaries are kept fixed in place until the next IR-sweep, and then removed.

%%%%%%%%%%%%%%%%%%
\begin{figure}
\includegraphics[width=8.8cm]{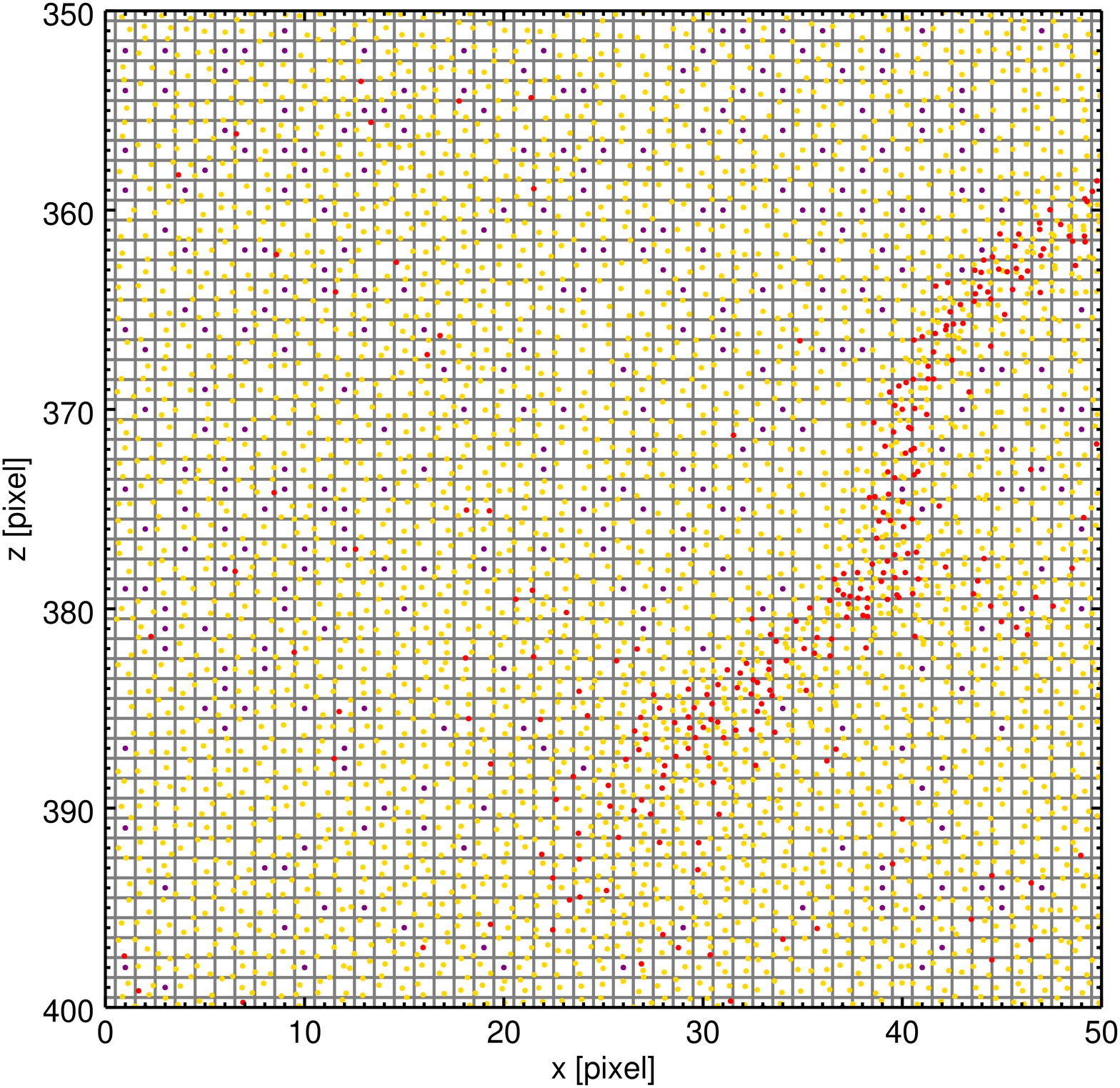}
\caption{Illustration of cork injection and removal. The figure shows a subfield of the same simulation as in Fig.~\ref{fig:void} at $t=300$~s after initialising with a homogenous cork distribution. The control volume was set to a single grid cell, the minimum cork density to 1 and the maximum cork density to 2. IR-sweeps were performed every 10~s. Corks marked for removal are red. Corks that have just been injected are purple, all other corks are yellow-gold. The grey lines indicate the borders of the control volumes. An animation showing cork removal and injection over a period of time is available online.}
\label{fig:showinject}
\end{figure}
%%%%%%%%%%%%%%%%%%

Figure~\ref{fig:showinject} and the accompanying animation illustrate the injection and removal process. The band with red corks is a shock wave passing through the domain. The cork density is above the maximum of 2 and all red corks will be removed to lower the cork density to the maximum. Purple corks are corks that will be injected to raise the cork density in each control volume to the minimum of 1. The yellow-gold corks are unaffected by the removal sweep.

The computing time associated with advecting the corks is negligible compared to the rest of the MHD scheme.

%%%%%%%%%%%%%%%%%%
\subsection{Construction of pathlines} \label{subsec:construct}
%%%%%%%%%%%%%%%%%%

Pathlines are constructed in a post processing step. During the simulation only the following is done: the simulation is initialised with corks in their starting positions. Each cork is given a unique ID (an integer), and their ID and positions are written to file. The corks are passively advected until an IR-sweeps starts. The sweeps are done in a particular order. First the injection sweep is performed and newly injected corks are assigned a new unique ID. Then all cork positions and IDs are written to file. I call this file a snapshot, labeled with $j$, and written out on simulation time $t_j$. Then the removal sweep is performed. This order is important in order to perform the construction of pathlines in the post processing step. 

The total number of unique corks that were present in the simulation is $N$, the IDs of the corks are integers, and we denote an ID as $i$.  The total number of snapshots is $j_\mathrm{tot} = j_\mathrm{start} - j_\mathrm{end}+1$. Because corks are injected and removed, not all corks exist in all snapshots. The pathline construction is done as follows:

\begin{itemize}
%\item All corks in each snapshot are put in order of increasing ID to increase the efficiency of the pathline construction.
\item An array is created that contains the snapshots where the corks are injected and removed ($g_i=j_\mathrm{injection}$ and $h_i=j_\mathrm{removal}$). This is a size $[2,N]$ array, which I call lifetime list ($\mathtt{LL}$), and $\mathtt{LL}[:,i] = (g_i, h_i)$.
\item For a given snapshot $j$, a list of corks of interest (COI) is created. Each cork has a different location, and is thus part of a unique pathline. The COIs, and thus the pathlines are labelled with $k, 1 \le k \le n_\mathrm{COI} = n_\mathrm{PL}$. 
\item A 2D-array called pathline ID ($\mathtt{PLID}$) is created of size ($n_\mathrm{PL}, j_\mathrm{tot}$). For each pathline $k$ we take the corresponding COI, with ID$_{k}=a$,  we look up $g_a$ and $h_a$ in the lifetime list and set $\mathtt{PLID}[k, g_a- j_\mathrm{start}:h_a-j_\mathrm{start}] = a$.
\item A loop forward in time over each snapshot  $f=j \ldots j_\mathrm{end}$ is performed, and then an inner loop over each pathline $k$. We look up whether cork $a$ still exists in snapshot $f+1$, if not the cork that is closest in space to cork $a$ in snapshot $f$ is taken (this closest cork has ID $b$) and set $\mathtt{PLID}[k, g_b- j_\mathrm{start}:h_b- j_\mathrm{start}] = b$. I.e., if a cork is destroyed we follow the path of the next closest cork to continue construction of the pathline. 
\item A loop backward in time over each snapshot  $f=j \ldots j_\mathrm{start},-1$ is performed, and then an inner loop over each pathline $k$. If cork $a$ exists in snapshot $f-1$ we continue to the next cork, if not the cork that is closest in space to cork $a$ in snapshot $f$ is taken (with ID $b$) and set $\mathtt{PLID}[k, g_b- j_\mathrm{start}:h_b-1- j_\mathrm{start}] = b$. I.e., if a cork is injected the path of the next closest cork is followed backward in time to continue construction of the pathline. 
\end{itemize}
The end result is that $\mathtt{PLID}[k,f]$ is the cork ID that follows pathline $k$ at the simulation time corresponding to snapshot $f$. From this list and the snapshots of coordinates of the corks one can create the pathline $\vec{r}_k(t)$. From this information any type of MHD or auxiliary variable along the pathline can be computed.

There is an asymmetry in the construction of pathlines forward in time and backward in time. In the forward construction a jump to a new cork can only occur in regions with high cork density (converging flow), and the jump in cork location is then guaranteed to be small: The average jump in units of grid cells is $n_\mathrm{max}^{-1/3}$, the maximum jump is the edge length of the control volume because there is always at least one cork per control volume. 

In the backward construction a jump to a new cork can only occur if the current cork was injected during the simulation. If only a single grid cell was empty then the average jump is the edge length of the control volume. However, if the the pathline traces through the centre of an area of strongly divergent flows then the jump can be multiple grid cells if the flow has expanded so much that it has removed corks from a region spanning multiple control volumes in the time interval between IR-sweeps. In the latter case this will show up as a large jump in the pathline coordinates. Inspection of Fig.~\ref{fig:showinject} shows that the latter is rarely the case at a snapshot interval $\Delta t =10$~s even in the case of the extreme expansion common in the chromosphere of weakly-magnetised 2D simulations 
\citep[cf.][]{2011A&A...530A.124L}.
%

%%%%%%%%%%%%%%%%%%
\begin{figure}

\includegraphics[width=8.8cm]{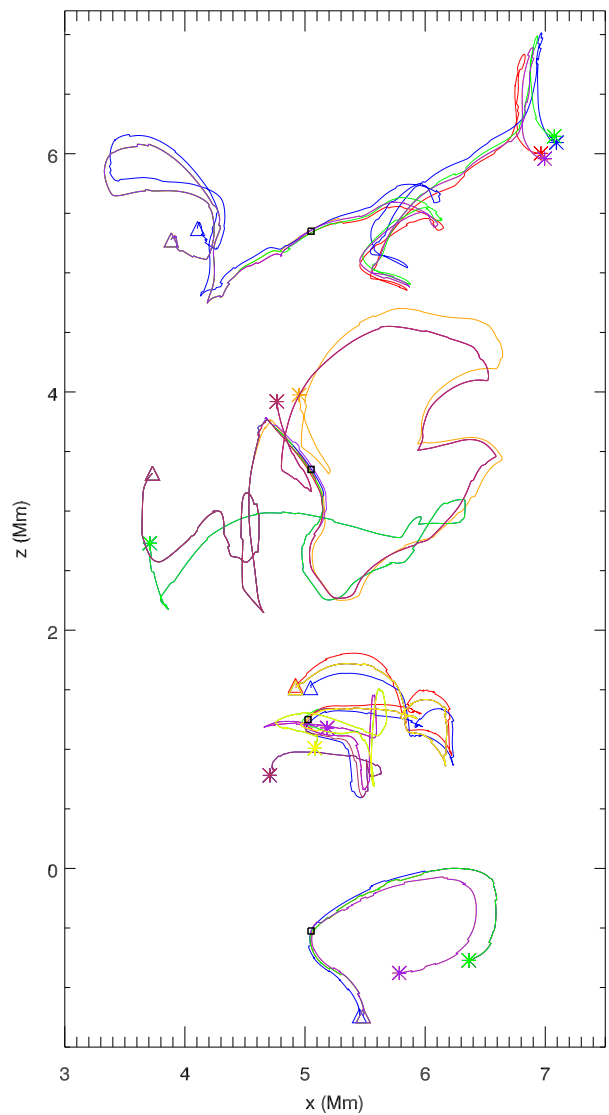}
\caption{Pathlines of closely spaced gas parcels. The coloured curves indicate individual pathlines, starting at $t=0$~s at the triangle, and ending at $t=1000$~s at the asterix. The pathlines are constructed from seed points at $t=500$~s located in the small black squares that have have a side length of 50~km, and are grouped into four areas. From top to bottom: corona, transition region, chromosphere, and convection zone.}
\label{fig:bunch}
\end{figure}
%%%%%%%%%%%%%%%%%%

Figure~\ref{fig:bunch} attempts to demonstrate errors made in pathline construction owing to the jump from one cork to the next. It shows pathlines constructed from seed points that are initially close (up to 1.5 grid cell edge) together. A drawback of the cork injection and removal process is that pathlines that are separate but close together at the seed points, can be merged. This happens in the forward-in-time construction if they both pass through the same grid cell in a compressive flow. Here the corks tracing the pathlines are removed and both pathlines jump to the closest remaining cork, which can be the same one. 

In the backward-in-time tracing this can happen if the corks traverse a region of expanding flows. If one, or both of the corks are injected then the pathlines might start following the same corks. A clear example in Figure~\ref{fig:bunch} is the transition region corks. Tracing the pathlines backward in time from the square to the triangle shows that the green, orange, and purple pathlines merge into a single pathline. 

Generally speaking most pathlines stay close together and show a similar trajectory for at least several hundred seconds. However, in case of large gradients in the atmosphere, a position difference of only a few grid cells can mean a large difference in atmospheric conditions. 

The errors of jumping from one cork to the next (which causes similar errors as selecting different pathlines from closely spaced pathlines)  are thus in general small, but can occasionally be large in regions with large gradients of atmospheric quantities. If the highest accuracy is required then using tracer particles without injection and removal is the better choice, but this comes at the price of an uneven sampling of the computational volume.

%%%%%%%%%%%%%%%%%%
\subsection{Simulation setup}
%%%%%%%%%%%%%%%%%%

I ran 2D simulations using Bifrost on a grid of $512\times496$ grid cells, spanning from 2.5~Mm below to 14~Mm above the height of average continuum optical depth unity. The simulation contained a very weak magnetic field with an average unsigned flux of only 2~G in the photosphere. This simulation does not generate enough dissipation of electric currents to maintain a corona. The upper boundary was therefore set to drive the temperature to 450~kK. The simulation is not particularly realistic, but it serves adequately to test the cork method, especially because a non-magnetic chromosphere shows very strong compression and expansion resulting from the acoustics shock generated by the convection. The method was also tested in 3D simulation, but because 3D results are more difficult to visualise without yielding extra insight in the method I show results from the 2D simulation only.

The simulations were initialised at $t=0$~s with a homogeneous cork distribution of one cork per grid cell. The simulations differ only in the time interval between IR-sweeps. Path lines were constructed between $t=0$~s and $t=1000$~s from COIs selected at $t=500$~s.

\section{Results} \label{sec:results}

\subsection{MHD equations in Lagrangian frame}

Bifrost solves the MHD equations in a Eulerian frame, with mass density $\rho$, velocity $\vu$, magnetic field $\vec{B}$ and internal energy per volume $e$ as fundamental variables. Rewriting these equations into Lagrangian evolution equations for the mass density , the velocity (which is equal to momentum per unit mass), and internal energy per unit mass $E$ yields
\bea
\ddt{\rho} & = & - \rho \dv \cdot \vu+ A_\mathrm{diff}, \label{eq:rho_evol} \\
\ddt{\vu} & = & - \frac{\dv P}{\rho} - \frac{\dv \cdot \tau}{\rho} + \frac{\vec{J} \times \vec{B}}{\rho} + \vec{g}+ B_\mathrm{diff}, \label{eq:u_evol}  \\
\ddt{E} & = & - \frac{P}{\rho} \dv \cdot \vu + \frac{Q}{\rho}+ C_\mathrm{diff}. \label{eq:E_evol}
\eea
Here, $\mathrm{d}/\mathrm{d}t$ is the co-moving derivative, $P$ is the gas pressure, $\tau$ the viscous stress tensor, $\vec{J}$ the current density, $\vec{g}$ the gravitational acceleration, and $Q$ energy gains and losses owing to non-adiabatic processes. The terms $A_\mathrm{diff}$, $B_\mathrm{diff}$, and $C_\mathrm{diff}$ represent artificial diffusion terms that are added to the equations in order to keep the numerical scheme stable. Explicit expressions for these terms can be found in 
\citet{2011A&A...531A.154G}.

The mass density changes as the fluid parcel expands or contracts (Eq.~\ref{eq:rho_evol}). A fluid parcel can be accelerated by gas pressure gradients, viscous shear forces, the Lorentz force, and gravity (Eq.~\ref{eq:u_evol}).  The internal energy changes through work done by compression and expansion and non-adiabatic energy losses and gains. In this case the latter includes Spitzer thermal conductivity, radiative losses and gains, and heating through Joule dissipation of electric currents (Eq.~\ref{eq:E_evol}). 

%%%%%%%%%%%%%%%%%%
\subsection{Example results}
%%%%%%%%%%%%%%%%%%

%%%%%%%%%%%%%%%%%%
\subsubsection{A convection cell pathline} \label{sec:granulePL}
%%%%%%%%%%%%%%%%%%

%%%%%%%%%%%%%%%%%%
\begin{figure*}
\centering
\includegraphics[width=17cm]{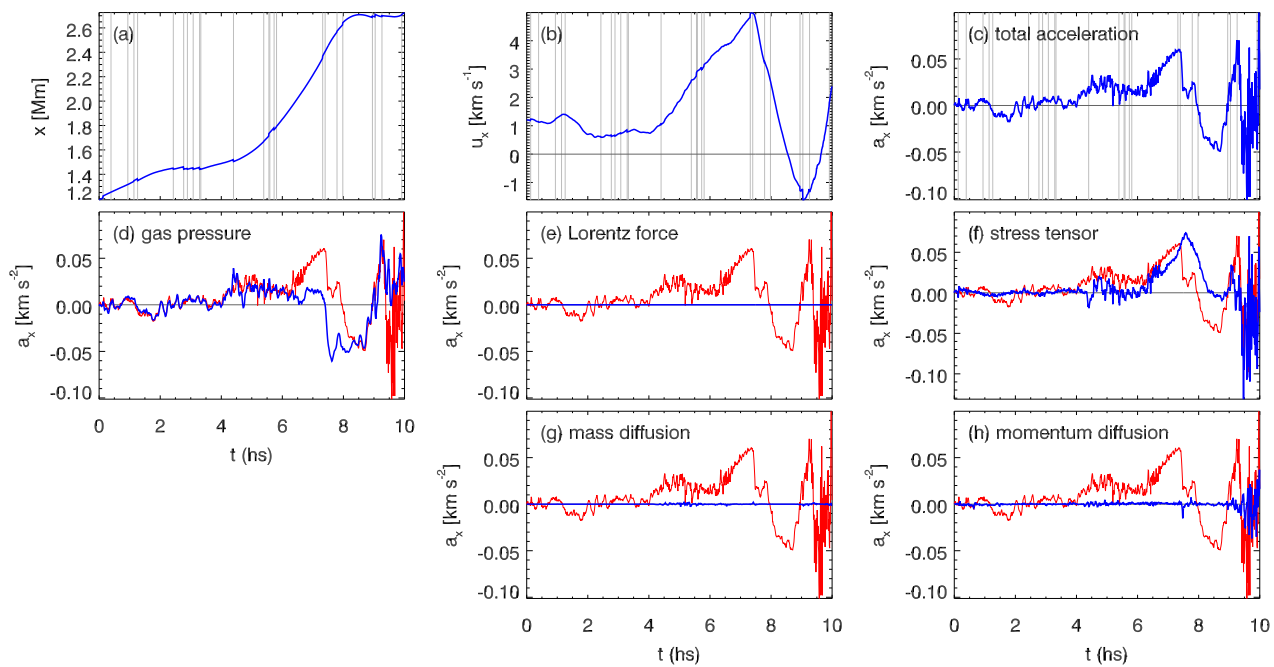}
\caption{Coordinate, speed, acceleration and forces acting in the $x$-direction on a fluid parcel in the convection zone as a function of time. Panels a, b and c show the coordinate, speed and acceleration. The thin vertical lines indicate the times where a different cork is used to trace the pathline. Panels d\,--\,h show the various forces in the simulation (blue) in comparison to the total acceleration (red). Panels g and h together make up the artificial diffusion term $B_\mathrm{diff}$ in Eq.~\ref{eq:u_evol}.}
\label{fig:force_x_125}
\end{figure*}
%%%%%%%%%%%%%%%%%%

%%%%%%%%%%%%%%%%%%
\begin{figure*}
\centering
\includegraphics[width=17cm]{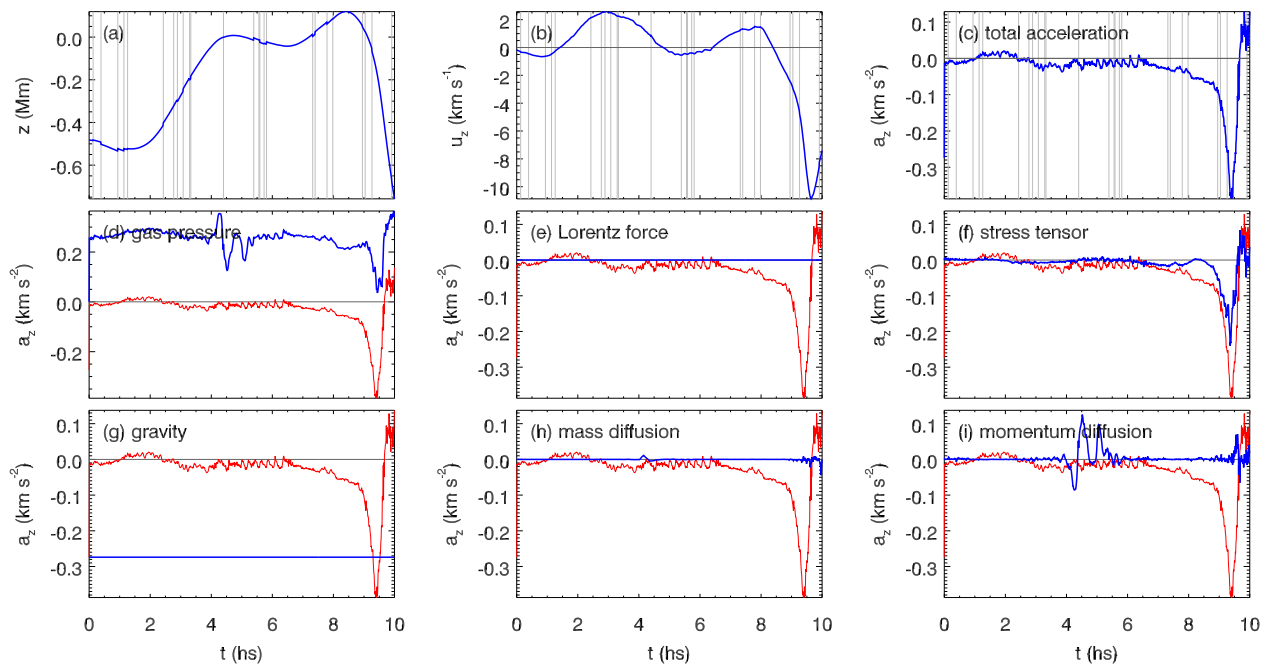}
\caption{As Fig.~\ref{fig:force_x_125}, but now for the $z$-direction. Panel~g shows the constant acceleration due to solar gravity.}
\label{fig:force_z_125}
\end{figure*}
%%%%%%%%%%%%%%%%%%

%%%%%%%%%%%%%%%%%%
\begin{figure*}
\centering
\includegraphics[width=17cm]{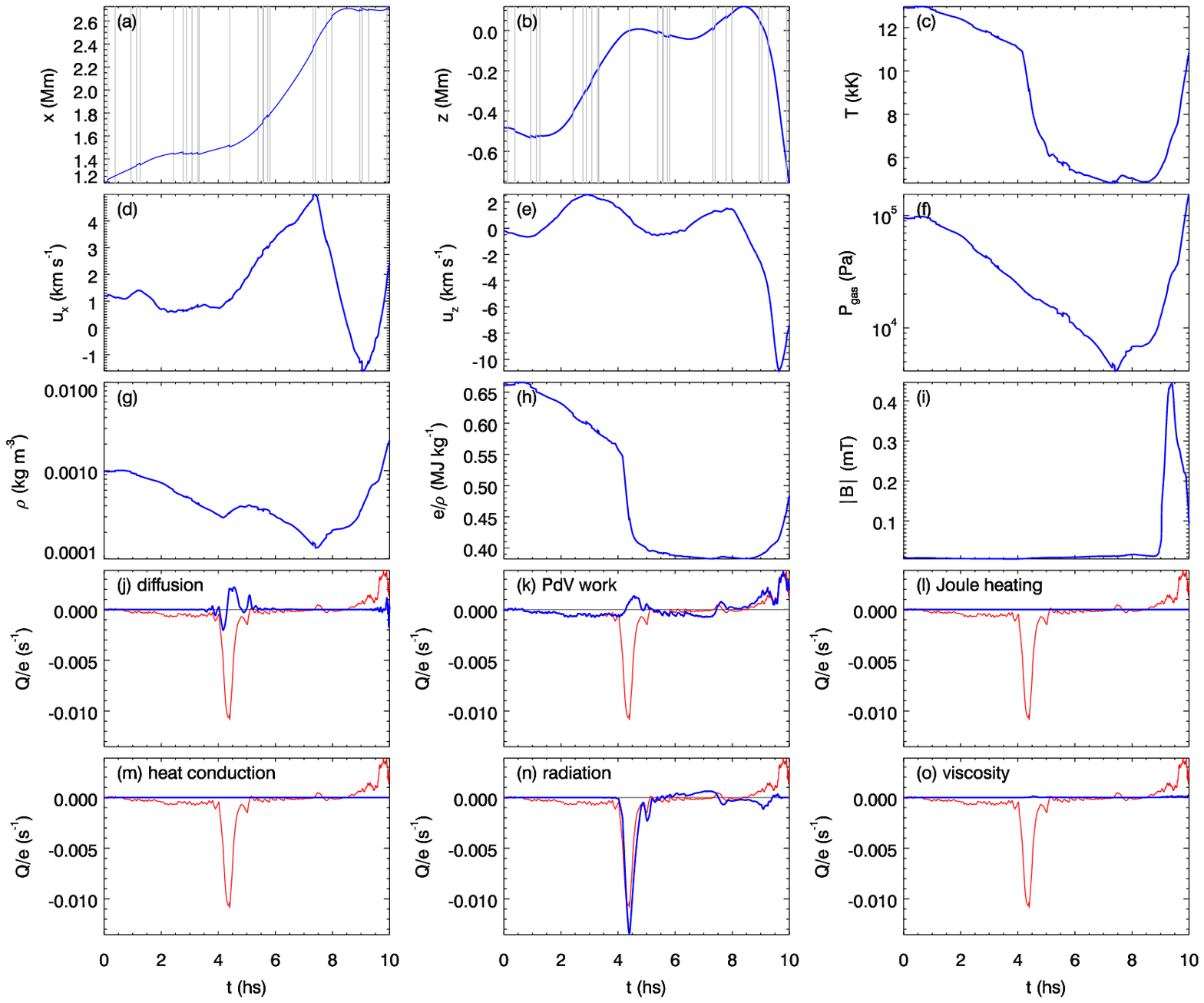}
\caption{Evolution of the convection-zone gas parcel shown in Figs.~\ref{fig:force_x_125} and~\ref{fig:force_z_125}. Quantities are indicated on the $y$-axes. Panels j\,--\,o show the heating and cooling terms, normalised to the total internal energy per volume. They thus indicate a relative rate of change per unit time. Panel j  shows the artificial diffusion term $C_\mathrm{diff}$ in Eq.~\ref{eq:E_evol}.}
\label{fig:evol_125}
\end{figure*}
%%%%%%%%%%%%%%%%%%

Figures~\ref{fig:force_x_125}\,--\,\ref{fig:evol_125} show the behaviour of a fluid parcel along a pathline in a convection cell by displaying the various terms in Eqs.~\ref{eq:rho_evol}--\ref{eq:E_evol}. The parcel rises to the photosphere, cools radiatively and sinks down again into the convection zone, a process that has been well studied and understood in earlier numerical simulations
\citep[e.g]{1982A&A...107....1N,1998ApJ...499..914S}. 
I show the results here as an illustrative case, as well as a test of the method
\citep[compare with Fig.~18 and Sec.~5.6 of ][]{1998ApJ...499..914S}.

The IR-sweep interval was 1~s, and the pathline was constructed from 25 individual corks. The transition from one cork to the next is indicated by the vertical grey lines in Figs~\ref{fig:force_x_125}\,--\,\ref{fig:evol_125}, and can also be seen as slight ($\sim 1$ grid cell) jumps in the coordinates of the pathline.

Figure~\ref{fig:force_z_125} shows the vertical motion of the parcel (panel a). It rises through the convection zone owing to gas pressure gradients, balanced by gravity (panels d and g). Around $t=400$~s radiative cooling sets in (Figure~\ref{fig:evol_125}n), the vertical pressure gradient drops and the parcel is decelerated. Figure~\ref{fig:force_x_125} show that around the same time the high pressure in the convection cell interior accelerates the parcel sideways to the edge of the cell (Panel b and d), where its horizontal motion is abruptly decelerated by pressure and stress forces around $t=800$~s (panels b, d, and f). At the same time the parcel  starts sinking down again into an intergranular lane (Figure~\ref{fig:force_z_125}b).

Figure~\ref{fig:evol_125} displays the evolution of some MHD quantities, as well as the internal energy of the same gas parcel. Of interest to note is the radiative cooling when the parcel reaches the surface, the radiative heating (by hot radiation from below) as it moves towards the edge of the convection cell and the second cooling episode when the parcel reaches the cell edge (panel n). 

%%%%%%%%%%%%%%%%%%
 \subsubsection{A  transition region pathline} \label{sec:TRPL}
%%%%%%%%%%%%%%%%%%

%%%%%%%%%%%%%%%%%%
\begin{figure*}
\centering
\includegraphics[width=17cm]{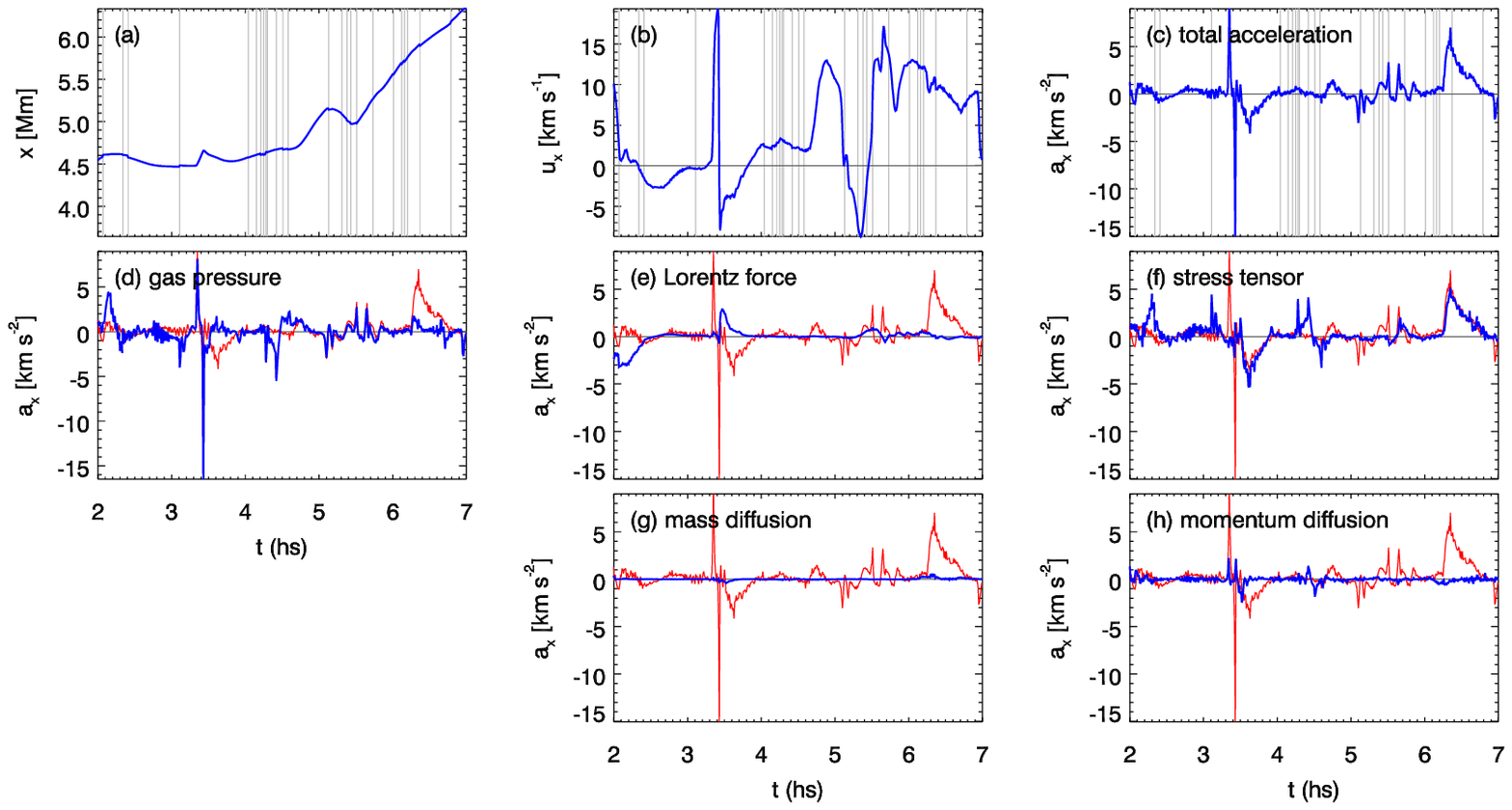}
\caption{As Fig.~\ref{fig:force_x_125}, but now for a gas parcel in the chromosphere and transition region.}
\label{fig:force_x_625}
\end{figure*}
%%%%%%%%%%%%%%%%%%

%%%%%%%%%%%%%%%%%%
\begin{figure*}
\centering
\includegraphics[width=17cm]{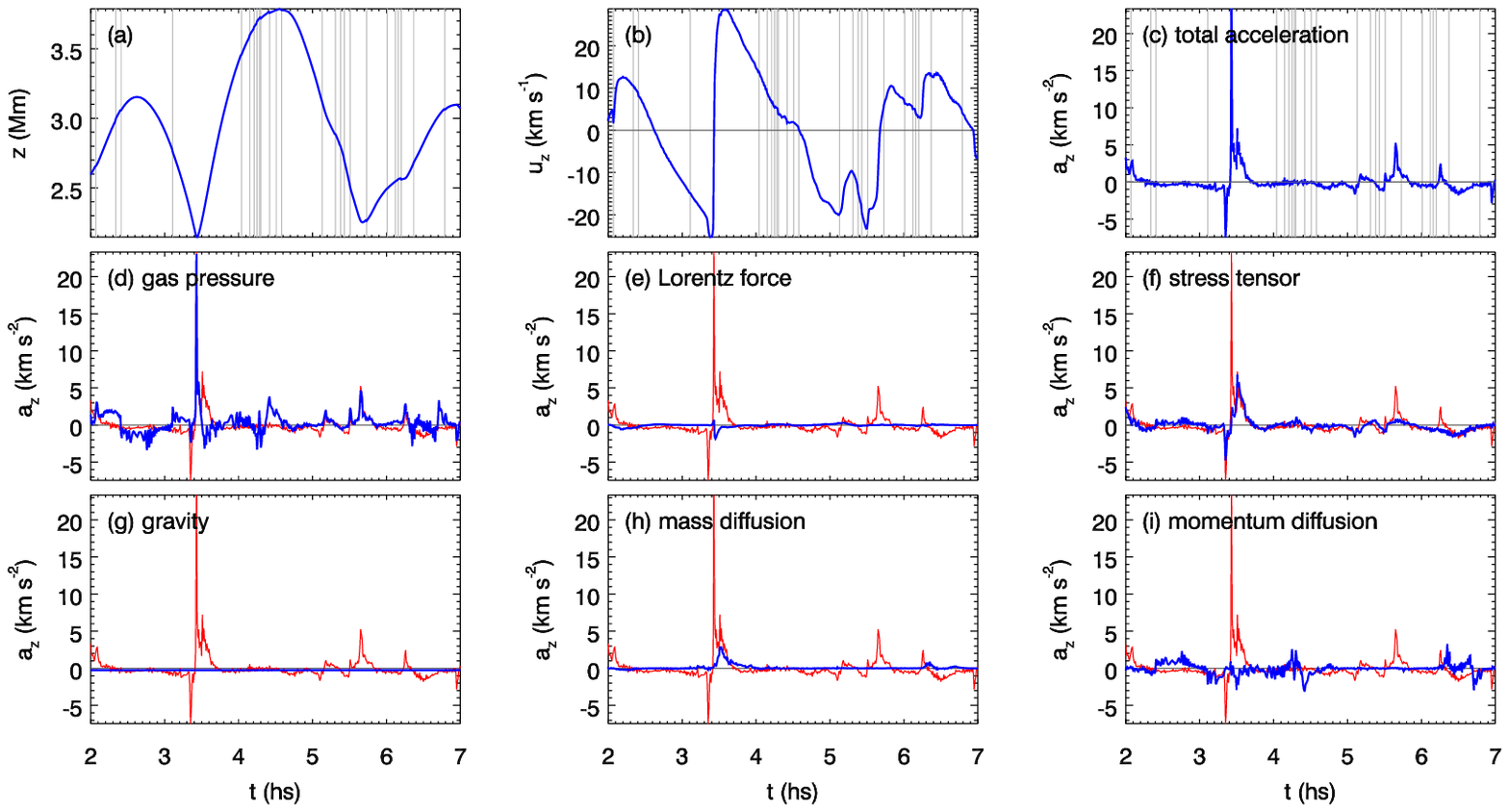}
\caption{As Fig.~\ref{fig:force_z_125}, but now for a gas parcel in the chromosphere and transition region.}
\label{fig:force_z_625}
\end{figure*}
%%%%%%%%%%%%%%%%%%

%%%%%%%%%%%%%%%%%%
\begin{figure*}
\centering
\includegraphics[width=17cm]{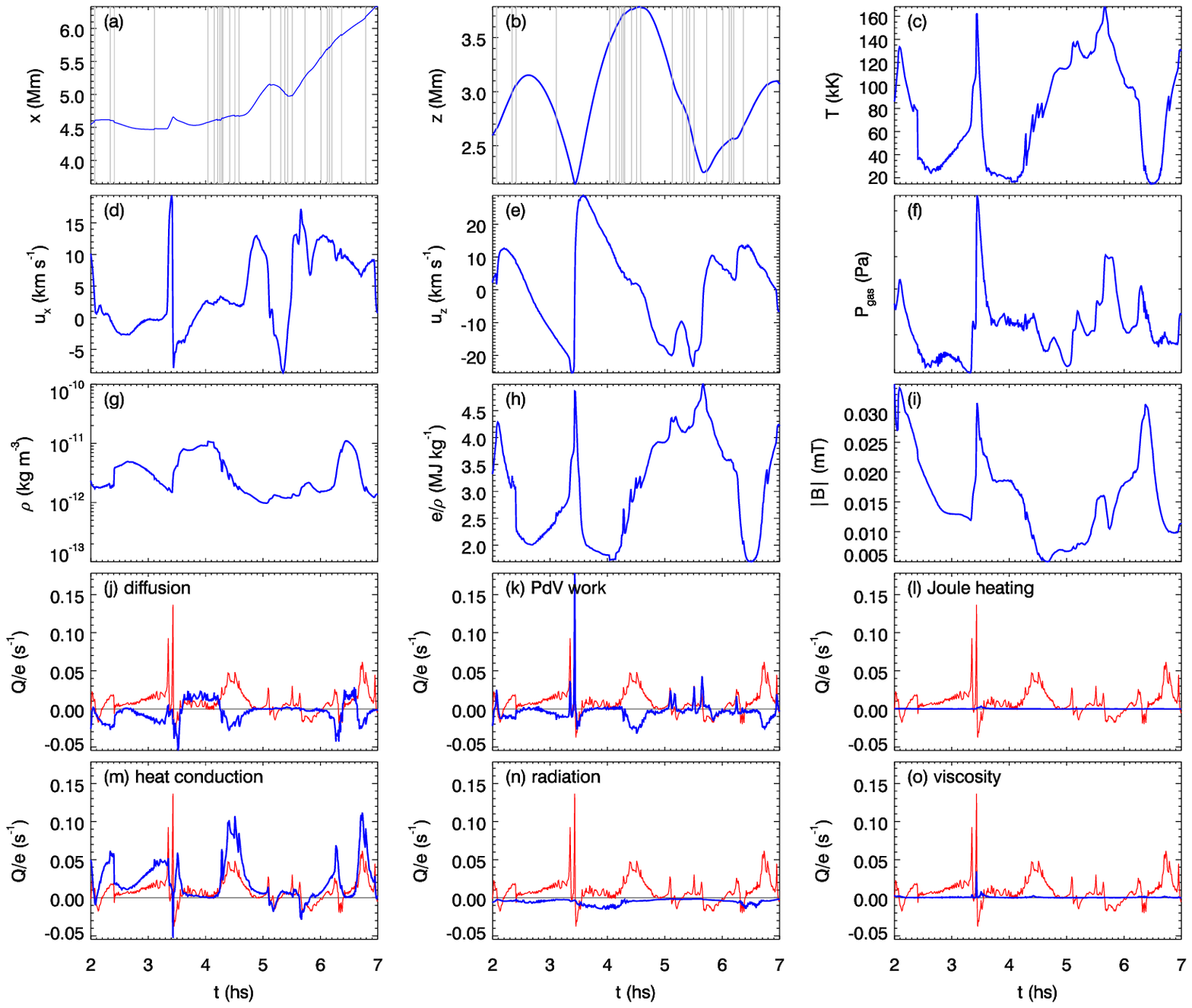}
\caption{As Fig.~\ref{fig:evol_125}, but now for a gas parcel in the chromosphere and transition region.}
\label{fig:evol_625}
\end{figure*}
%%%%%%%%%%%%%%%%%%

Figures~\ref{fig:force_x_625}\,--\,\ref{fig:evol_625} show the behaviour of a fluid parcel along a pathline located in the upper chromosphere and transition region, from the same simulation as the cork described in Sec.~\ref{sec:granulePL}. The dynamics of the parcel are dominated by acoustic waves and heat conduction. The simulation is only weakly magnetised, so that the Lorentz force is of minor importance (Figures~\ref{fig:force_x_625}e and~\ref{fig:force_z_625}e). The velocities, forces, and compression and expansion rate are much larger than for the convection zone pathline. Many more injections and removals are thus taking place, and the pathline is constructed from 46 corks, compared to 25 in the convection zone pathline. I only show a time period of 500~s so that the rapid changes are better visible.

The acceleration along the pathline is very bursty, with brief spikes of acceleration caused by gas pressure gradients when the pathline crosses the shock waves (Figures~\ref{fig:force_x_625}d and~\ref{fig:force_z_625}d). The spikes typically last between 2~s and 5~s, which is consistent with the time it takes for the pathline to cross 2-3 grid cells (the thickness of the shock front in the code) at a relative speed of about 20~\kms. The viscous stresses dominate the acceleration that occurs on slightly longer timescales than the fast shock acceleration (Figures~\ref{fig:force_x_625}f and~\ref{fig:force_z_625}f).

Figure.~\ref{fig:evol_625} shows the evolution of the energy balance along the pathline. The temperature varies strongly between <20~kK to more than 100~kK with a period of roughly 3~min. Panel c, g, and h show distinct jumps between $t=400$~s and  $t=460$~s. Inspection of the vertical lines in panels a and b show that these jumps are unphysical and instead associated with jumps from one cork to the next. Panel g shows that this occurs in a strongly expanding region, and it occurs before $t=500$~s, i.e, while tracing the pathline backward in time from its seed location and time (see the last paragraph of Sec.~\ref{subsec:construct}). 

The variations in temperature are largely driven by the heat conduction, only at t=340~s is compressive heating clearly dominant. Another interesting behaviour is that the non-physical energy diffusion (panel j) has a somewhat smaller magnitude than the thermal conductivity, but tends to have an opposite sign. The fluid parcel is constantly losing energy through radiation (Figure.~\ref{fig:evol_625}n). 

%%%%%%%%%%%%%%%%%%
\section{Summary and discussion} \label{sec:summary}
%%%%%%%%%%%%%%%%%%

In this paper I presented a method to trace flows in radiation-MHD simulations of compressive flows in the Lagrangian frame, computed with the same accuracy as the MHD variables, which  themselves are solved on a Eulerian grid. The method is based on passively-advected tracer particles (corks). 
The key observation is that MHD variables vary only slowly on spatial scales of the order of a grid cell. Errors in tracing path lines of the order of a grid cell are thus often acceptable. 
By periodically injecting and removing corks the method keeps the particle-density close to one cork per grid cell at all times and locations in the computational domain. In a post-processing steps pathlines can be traced forward and backward in time from any location and any time by stitching together individual cork tracks. 

Errors from switching corks are typically small, but not negligible, especially in the transition region. The method also suffers from the merging of distinct pathlines into a single one in the case of strongly converging or expanding flows. 

The method allows tackling problems where the evolution of individual gas parcels is important, such as mass dredge-up during flux emergence, mass loading of chromospheric fibrils, mass and energy cycling between the chromosphere and the corona, and flow acceleration during reconnection. In regions where the frozen-in approximation holds the tracer particles could also be used to accurately follow the evolution of magnetic field lines in time 
\citep[e.g.,][]{2013ApJ...771...20M,2015ApJ...802..136L}. 
By tracing how corks that are initially located on the same field line drift to different field lines over time, magnetic diffusion and reconnection could also be studied.

%%%%%%%%%%%%%
\begin{acknowledgements}
 Some computations were performed on resources provided by the Swedish National
 Infrastructure for Computing (SNIC) at the PDC Centre for High Performance Computing (PDC-HPC)
 at the Royal Institute of Technology in Stockholm as well at the High
 Performance Computing Center North (HPC2N). JL has received support through grants from the Swedish
 Research  Council (2017-04099) and the Knut och och Alice Wallenberg foundation (2016.0019).
 \end{acknowledgements}
%%%%%%%%%%%%%

\bibliographystyle{aa} % style aa.bst
\bibliography{corks}

\begin{thebibliography}{15}
\expandafter\ifx\csname natexlab\endcsname\relax\def\natexlab#1{#1}\fi

\bibitem[{{Cheung} {et~al.}(2007{\natexlab{a}}){Cheung}, {Sch{\"u}ssler}, \&
  {Moreno-Insertis}}]{2007A&A...467..703C}
{Cheung}, M.~C.~M., {Sch{\"u}ssler}, M., \& {Moreno-Insertis}, F.
  2007{\natexlab{a}}, \aap, 467, 703

\bibitem[{{Cheung} {et~al.}(2007{\natexlab{b}}){Cheung}, {Sch{\"u}ssler}, \&
  {Moreno-Insertis}}]{2007A&A...461.1163C}
{Cheung}, M.~C.~M., {Sch{\"u}ssler}, M., \& {Moreno-Insertis}, F.
  2007{\natexlab{b}}, \aap, 461, 1163

\bibitem[{{Freytag} {et~al.}(2012){Freytag}, {Steffen}, {Ludwig},
  {Wedemeyer-B{\"o}hm}, {Schaffenberger}, \& {Steiner}}]{2012JCoPh.231..919F}
{Freytag}, B., {Steffen}, M., {Ludwig}, H.-G., {et~al.} 2012, Journal of
  Computational Physics, 231, 919

\bibitem[{{Gudiksen} {et~al.}(2011){Gudiksen}, {Carlsson}, {Hansteen}, {Hayek},
  {Leenaarts}, \& {Mart{\'{\i}}nez-Sykora}}]{2011A&A...531A.154G}
{Gudiksen}, B.~V., {Carlsson}, M., {Hansteen}, V.~H., {et~al.} 2011, \aap, 531,
  A154

\bibitem[{{Leenaarts} {et~al.}(2011){Leenaarts}, {Carlsson}, {Hansteen}, \&
  {Gudiksen}}]{2011A&A...530A.124L}
{Leenaarts}, J., {Carlsson}, M., {Hansteen}, V., \& {Gudiksen}, B.~V. 2011,
  \aap, 530, A124

\bibitem[{{Leenaarts} {et~al.}(2015){Leenaarts}, {Carlsson}, \& {Rouppe van der
  Voort}}]{2015ApJ...802..136L}
{Leenaarts}, J., {Carlsson}, M., \& {Rouppe van der Voort}, L. 2015, \apj, 802,
  136

\bibitem[{{Moreno-Insertis} \& {Galsgaard}(2013)}]{2013ApJ...771...20M}
{Moreno-Insertis}, F. \& {Galsgaard}, K. 2013, \apj, 771, 20

\bibitem[{{N{\'o}brega-Siverio} {et~al.}(2016){N{\'o}brega-Siverio},
  {Moreno-Insertis}, \& {Mart{\'{\i}}nez-Sykora}}]{2016ApJ...822...18N}
{N{\'o}brega-Siverio}, D., {Moreno-Insertis}, F., \& {Mart{\'{\i}}nez-Sykora},
  J. 2016, \apj, 822, 18

\bibitem[{{Nordlund}(1982)}]{1982A&A...107....1N}
{Nordlund}, A. 1982, \aap, 107, 1

\bibitem[{{Shelyag} {et~al.}(2013){Shelyag}, {Cally}, {Reid}, \&
  {Mathioudakis}}]{2013ApJ...776L...4S}
{Shelyag}, S., {Cally}, P.~S., {Reid}, A., \& {Mathioudakis}, M. 2013, \apjl,
  776, L4

\bibitem[{{Stein} \& {Nordlund}(1998)}]{1998ApJ...499..914S}
{Stein}, R.~F. \& {Nordlund}, {\AA}. 1998, \apj, 499, 914

\bibitem[{{Tortosa-Andreu} \& {Moreno-Insertis}(2009)}]{2009A&A...507..949T}
{Tortosa-Andreu}, A. \& {Moreno-Insertis}, F. 2009, \aap, 507, 949

\bibitem[{{V{\"o}gler} {et~al.}(2005){V{\"o}gler}, {Shelyag}, {Sch{\"u}ssler},
  {Cattaneo}, {Emonet}, \& {Linde}}]{2005A&A...429..335V}
{V{\"o}gler}, A., {Shelyag}, S., {Sch{\"u}ssler}, M., {et~al.} 2005, \aap, 429,
  335

\bibitem[{{Wedemeyer} \& {Steiner}(2014)}]{2014PASJ...66S..10W}
{Wedemeyer}, S. \& {Steiner}, O. 2014, \pasj, 66, S10

\bibitem[{{Wray} {et~al.}(2015){Wray}, {Bensassi}, {Kitiashvili}, {Mansour}, \&
  {Kosovichev}}]{2015arXiv150707999W}
{Wray}, A.~A., {Bensassi}, K., {Kitiashvili}, I.~N., {Mansour}, N.~N., \&
  {Kosovichev}, A.~G. 2015, ArXiv e-prints

\end{thebibliography}

\end{document}